\documentclass[a4paper]{article}

\usepackage{INTERSPEECH2019, graphicx, amssymb, epsfig, cite, fixmath, subfigure}
\usepackage{multirow}

\title{An End-to-End Text-independent Speaker Verification Framework with a Keyword Adversarial Network}
\name{Sungrack Yun, Janghoon Cho, Jungyun Eum, Wonil Chang, Kyuwoong Hwang}
\address{
  Qualcomm AI Research${}^{\dagger}$, Qualcomm Korea YH \thanks{  ${}^{\dagger}$ Qualcomm AI Research is an initiative of Qualcomm Technologies, Inc.}}
\email{\{sungrack, janghoon, c\_jeum, wichang, kyuwoong\}@qti.qualcomm.com}

\begin{document}

\maketitle
\begin{abstract}
  This paper presents an end-to-end text-independent speaker verification framework by jointly considering the speaker embedding (SE) network and automatic speech recognition (ASR) network.
  The SE network learns to output an embedding vector which distinguishes the speaker characteristics of the input utterance, while the ASR network learns to recognize the phonetic context of the input. In training our speaker verification framework, 
  we consider both the triplet loss minimization and adversarial gradient of the ASR network to obtain more discriminative and text-independent speaker embedding vectors. With the triplet loss, the distances between the embedding vectors of the same speaker are minimized while those of different speakers are maximized. Also, with the adversarial gradient of the ASR network, the text-dependency of the speaker embedding vector can be reduced. In the experiments, we evaluated our speaker verification framework using the LibriSpeech and CHiME 2013 dataset, and the evaluation results show that our speaker verification framework shows lower equal error rate and better text-independency compared to the other approaches.

\end{abstract}

\section{Introduction}
\label{sec:intro}

With the increasing number of researches, developments, and improvements on automatic speech recognition (ASR) \cite{audhkhasi2017direct, li2017acoustic, prabhavalkar2017comparison, zhao2018domain}, speaker verification (SV) \cite{heigold2016end, song2018noise, snyder2017deep, wan2018generalized, zhang2017end, zhong2017dnn}, and spoken dialog system \cite{liu2017iterative, ranzenberger2018integration}, the voice interface has been widely adopted in various artificial intelligent (AI) applications such as mobile phones, smart home IoT devices, and automotive infotainment system. 
Especially, the speaker verification and recognition have been crucial components in several AI speakers \cite{li2017acoustic, lopez2017alexa, purington2017alexa, zhao2018domain} for user authentication and personalized responses: given the user's voice command, the AI speaker can identify the user's voice and provide a user-specific services, e.g. music recommendation, equalizer adjustment, or schedule notification.

A number of researches on speaker verification and recognition have been proposed \cite{song2018noise, heigold2016end, wan2018generalized, zhong2017dnn, snyder2018x, zhang2017end}. In \cite{zhang2017end}, the authors presented an end-to-end text-independent speaker verification for variable utterance length by applying a spatial pyramid pooling layer to the inception-resnet architecture.
In \cite{song2018noise}, a simple pre-processing method to select noise-invariant frames from utterance was proposed for the text-independent SV system in unknown noisy environments. 
In \cite{heigold2016end, wan2018generalized}, the tuple-based end-to-end (TE2E) loss and generalized end-to-end (GE2E) loss were introduced to improve speaker verification models. 
In \cite{zhong2017dnn}, a deep neural network (DNN) was used to extract phonetically-aware i-vector and also bottleneck feature for short, text-constrained utterances. 
In \cite{snyder2018x}, a data augmentation technique was investigated to improve the speaker recognition performance with the DNN-based embeddings trained to discriminate speakers.   

\begin{figure}[t]
	\centering
	\centerline{\epsfig{figure=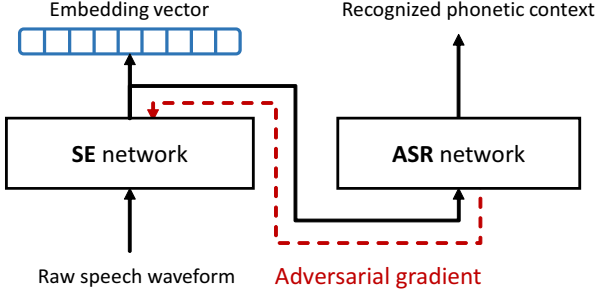,height=4.3cm}}
	\caption{Block diagram of the proposed SV framework: adversarial gradient from the ASR is used to obtain text-independent speaker embedding vector. The SE network can verify the user's voice independently of text-phrases.
	} \label{fig:advs_training}
\end{figure}

This paper presents an end-to-end text-independent SV framework by jointly considering two components: the speaker embedding (SE) network and the ASR network.
It can be considered that the ASR and SV are inversely-related. 
The ASR network distinguishes and classifies the phonetic context of the input utterances (text-dependent) from any speakers (speaker-independent), while the SE network extracts the speaker's identity (speaker-dependent) regardless of the input text-phrases (text-independent). 
Inspired from this property, we propose a text-independent SV framework where the SE network is combined jointly with the ASR network, as illustrated in Fig. \ref{fig:advs_training}. 
The SE network takes the raw speech waveform as input and outputs the speaker embedding vector using a deep end-to-end architecture consisting of a number of residual blocks \cite{cai2018deep, he2016deep, heZRS16}, convolution layers, and an attention layer. The ASR network classifies the phonetic context of the embedding vector, and the adversarial gradient \cite{domain_advs, goodfellow2014generative, pascual2017segan, radford2015unsupervised} is applied to the SE network such that the embedding vector is trained to be text-independent. Although we may also apply the adversarial gradient from the SE network to the ASR network for the speaker-independency in ASR, we only focus on the text-independent SE network in this research. In training SE network, we also combine the triplet loss \cite{zhang2017end} together with the adversarial gradient to obtain more discriminative speaker embedding vectors: the distances between the embedding vectors of the same speaker are minimized while those of different speakers are maximized.

The proposed SV framework was evaluated using the LibriSpeech \cite{panayotov2015librispeech} and CHiME 2013 dataset \cite{vincent2013second}. In the evaluation results, our SV framework shows lower equal error rate (EER) and better text-independent property compared to the other approaches. 

\section{Speaker Verification}
\label{sec:SV}

Speaker verification is a decision process of accepting or rejecting an input utterance $\bf x$ based on the speaker characteristics, and it can be accomplished by comparing $\bf x$ with the reference speaker model ${\bf X}_{ref}$ as:
\begin{eqnarray}
f({\bf X}_{ref}, {\bf x}) \underset{reject}{\overset{accept}{\gtrless}} \tau
\end{eqnarray}
where $f(\cdot, \cdot)$ measures the similarity score between ${\bf X}_{ref}$ and $\bf x$. 
If the score is greater than a pre-defined threshold $\tau$, $\bf x$ is accepted as a reference speaker's utterance; otherwise, it is rejected. 
The input observation $\bf x$ can be a raw speech waveform itself or an encoded vector using various feature extraction algorithms for speaker verification such as Mel-frequency cepstral coefficients (MFCCs) \cite{anguera2012speaker}, i-vector \cite{shum2013, dehak2011front, sell2014speaker, woubie2016improving}, or speaker embedding vectors \cite{snyder2017deep, snyder2018x, heigold2016end, wan2018generalized}. 
In this paper, we model the raw speech waveform directly and extract a $D$-dimensional speaker embedding vector for ${\bf x}$. 
The reference speaker model ${\bf X}_{ref}$ contains $M$ enrollment embedding vectors ${\bf X}_{ref} = \{ {\bf x}_{e_1}, ..., {\bf x}_{e_M} \}$, and we define the score function $f(\cdot, \cdot)$ based on the cosine similarity:
\begin{eqnarray}
f({\bf X}_{ref}, {\bf x}) = { { {\bf x} \cdot {\bf x}_{e}^{cent} } \over { ||{\bf x}|| ~ || {\bf x}_{e}^{cent} ||} }
\end{eqnarray}
where ${\bf x}_{e}^{cent} = {\sum_{i=1}^M {\bf x}_{e_i} }/M$.

\begin{figure}[t] 
	\begin{minipage}[b]{1.0\linewidth} 
		\centering
		\centerline{\epsfig{figure=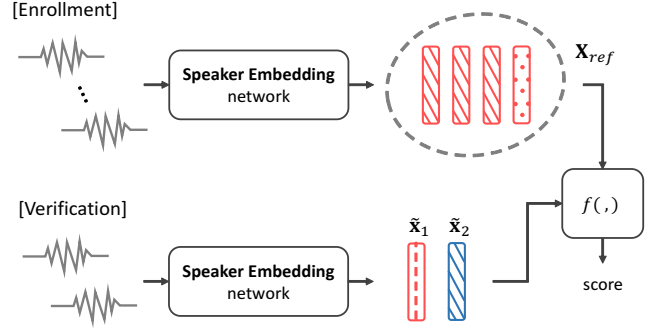,height=4.5cm}}
	\end{minipage}
	\caption{The process of speaker model enrollment and test vector verification. Color (red, blue) and texture (diagonal, dotted, dashed) represent speaker and text phrase, respectively.
	} \label{fig:SV_enroll_verification}
\end{figure}

An example of speaker model enrollment and test vector verification is illustrated in Fig. \ref{fig:SV_enroll_verification}: the color and texture of the embedding vectors represent the speaker and phonetic context, respectively. 
Here, we assume that the reference speaker model has the enrollment utterance of two different text phrases (diagonal and dotted). 
The test vector of the reference speaker, $\widetilde{\bf x}_1$, is from an unseen text phrase (dashed), and the test vector of an imposter, $\widetilde{\bf x}_2$, is from the seen text phrase (diagonal). 
Ideally, in the text-independent SV, $\widetilde{\bf x}_1$ should be accepted independently of the text phrase. 
In real-cases, however, $\widetilde{\bf x}_1$ may be falsely rejected due to its low score since ${\bf X}_{ref}$ does not contain the enrollment vector of the phrase (dashed). 
On the contrary, $\widetilde{\bf x}_2$ may be falsely accepted because the same phrase utterance (diagonal) exists in the enrollment set.
As described in this example, the text-independent SV may have a performance degradation when the reference speaker model is enrolled with the vectors from few specific text-phrases. 
This may often happen in real-applications: a speaker model is enrolled with one or two voice commands, and the user speaks the other voice commands which need to be verified. In the next section, we will describe our text-independent SV framework where the embedding vector is extracted from a deep end-to-end neural net architecture with an adversarial ASR network.

\section{Proposed SV Framework}
\label{sec:proposed_SV}

As illustrated in Fig \ref{fig:advs_training}, the proposed SV framework consists of 
two components: the SE network and the ASR network.
The SE network takes the raw speech waveform as the input and outputs an embedding vector, and the ASR network takes the embedding vector as the input and outputs the recognized phonetic context. In training the SE network, we use the adversarial gradient from the ASR to encourage the SE network to extract the embedding vector which is phonetically-independent and contains only speaker characteristics of the input.

\subsection{Deep End-to-End SE Network}
\label{ssec:deep_e2e}

\begin{figure}[t] 
	\begin{minipage}[b]{1.0\linewidth} 
		\centering
		\centerline{\epsfig{figure=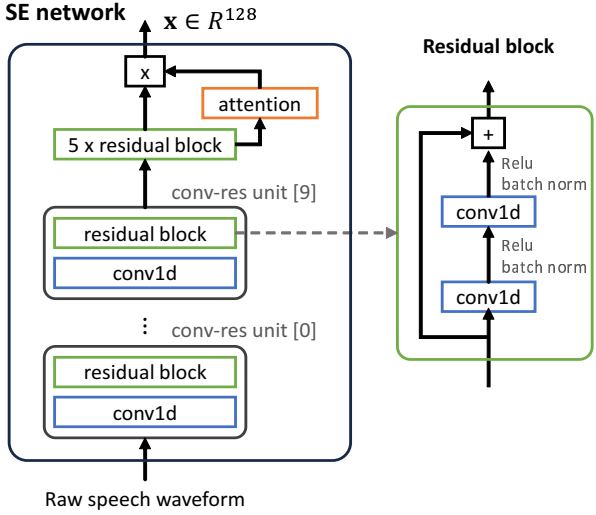,width=8.4cm}}
	\end{minipage}
	\caption{The proposed end-to-end SE network: 10 conv-res units, 5 residual blocks, and one attention layer are used to extract the embedding vector $\bf x$ from raw speech waveform. } \label{fig:deep_e2e_res}
\end{figure}

As illustrated in Fig. \ref{fig:deep_e2e_res}, the proposed end-to-end SE network extracts 128-dimensional embedding vector from raw speech waveform with the architecture of 10 conv-res units, 5 residual blocks \cite{cai2018deep, he2016deep, heZRS16}, and one attention layer. Each conv-res unit consists of 1-dimensional convolution layer and residual block. From the first to seventh conv-res unit, the number of input-output channels in convolution is doubled (i.e. 1$\rightarrow$2$\rightarrow\cdots\rightarrow$128) while it keeps the same number for remaining 3 conv-res units. The convolution width of the first conv-res unit is the length of the input waveform, and it is reduced by half at every conv-res unit (from the first to tenth). The residual block, illustrated in Fig. \ref{fig:deep_e2e_res}, consists of two convolution layers with batch normalization \cite{ioffeS15} and Relu, and a shortcut is connected from the input to output. The output of 5 residual blocks is combined with an attention layer to focus on more relevant region of the input, and finally we obtain the 128-dimensional embedding vector by averaging over the width (time-average).




\subsection{Training Loss for SE Network}
\label{ssec:kwd_adv_loss}


In training the SE network, we use the triplet loss and the adversarial gradient from the ASR network to obtain more discriminative and text-independent speaker embedding vectors. 
\subsubsection{Triplet loss}
\begin{figure}[t]
	\centering
	\centerline{\epsfig{figure=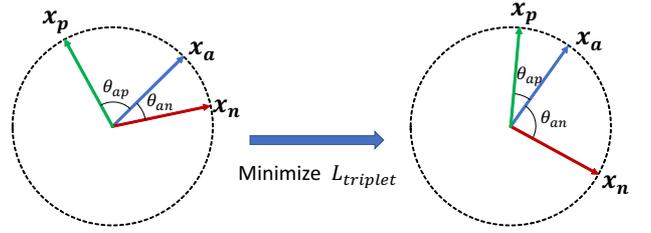,height=3.2cm}}
	\caption{Training with the triplet loss pulls the positive vector towards the anchor while pushes the negative vector away from the anchor.
	} \label{fig:triplet}
\end{figure}

The objective of the triplet loss is to maximize the similarity between the embedding vectors of the same speakers while minimize that of different speakers:
\begin{eqnarray}\label{eq:triplet}
f({{\bf x}_a}, {{\bf x}_p}) - \delta > f({{\bf x}_a}, {{\bf x}_n})
\end{eqnarray}
where $\delta$ quantifies the minimum margin between two similarities. Given the anchor ${{\bf x}_a}$, the positive sample ${{\bf x}_p}$ is selected from the same speaker with the anchor while the negative ${{\bf x}_n}$ is selected from the different speaker.
With the cosine similarity and normalized embedding vectors, the inequality (\ref{eq:triplet}) becomes 
\begin{eqnarray}\label{eq:triplet_angle}
	\cos{\theta_{ap}} > \cos{\theta_{an}} + \delta
\end{eqnarray}
where $\theta_{ap}$ and $\theta_{an}$ are respectively the angle between ${{\bf x}_a}$ and ${{\bf x}_p}$, and the angle between ${{\bf x}_a}$ and ${{\bf x}_n}$. 
As illustrated in Fig. \ref{fig:triplet}, the minimization of the triplet loss pulls together the anchor and positive vector, while pushes apart the negative from the anchor vector. In the training, we can only choose the triplets which violate the condition in (\ref{eq:triplet}), and the loss can be expressed as 
\begin{equation}\label{eq:triplet_final}
	L_{triplet} = -\min{ \left(   \cos \theta_{ap} - \cos \theta_{an}   , \delta \right)}.
\end{equation}

\subsubsection{Adversarial training loss from the ASR}



As illustrated in Fig. \ref{fig:advs_training}, the ASR network is combined with the SE network, and the adversarial gradient from the ASR is used to train the SE network for the text-independent speaker embedding vector. The objective of the ASR network is to accurately recognize the phonetic context of the embedding vector, and the network can be trained by minimizing the cross entropy between the true target label $y$ and the recognized label $\hat y$ (output of the ASR network): 
\begin{eqnarray}
L_{ASR} = - \sum_y  y \log ({\hat y}) 
\end{eqnarray}
where the recognition unit $y$ can be a character, phoneme or word. In this research, 
we use the word unit for $y$, and a DNN is used for the ASR network where the final soft-max layer gives the classification decision of $N$ keywords. Since the ASR is used {\it adversarially} in training the SE network, we can obtain the embedding vector which does not discriminate the word difference, i.e. text-independent.



Combining the triplet loss in Eq. (\ref{eq:triplet_final}) with the adversarial training loss from the ASR, we can obtain the entire loss to train the SE network:
\begin{eqnarray}
L_{SE} = L_{triplet} - \gamma L_{ASR} 
\end{eqnarray}
where $\gamma$ is a pre-determined value which controls the adversarial factor.

\section{Experiments}
\label{sec:exp}

\subsection{Dataset and Training}
\label{subsec:data}

The proposed SV framework is trained by two stages: training the baseline SE network and fine-tuning the SE network using the triplet loss combined with the ASR-adversarial loss. In training the baseline SE network, we used the LibriSpeech dataset \cite{panayotov2015librispeech} which contains 1,000 hours of 2,400 speakers' recordings based on the text from Project Gutenberg \cite{guten}. First, we randomly segmented the LibriSpeech utterances into the audio samples of a length between 1.5 and 2.0 sec. And, given these segmented audio samples, we trained the baseline SE network with a final softmax layer which is designed to classify 2,400 speakers by minimizing the cross entropy loss.

\begin{figure}[b]
	\centering
	\centerline{\epsfig{figure=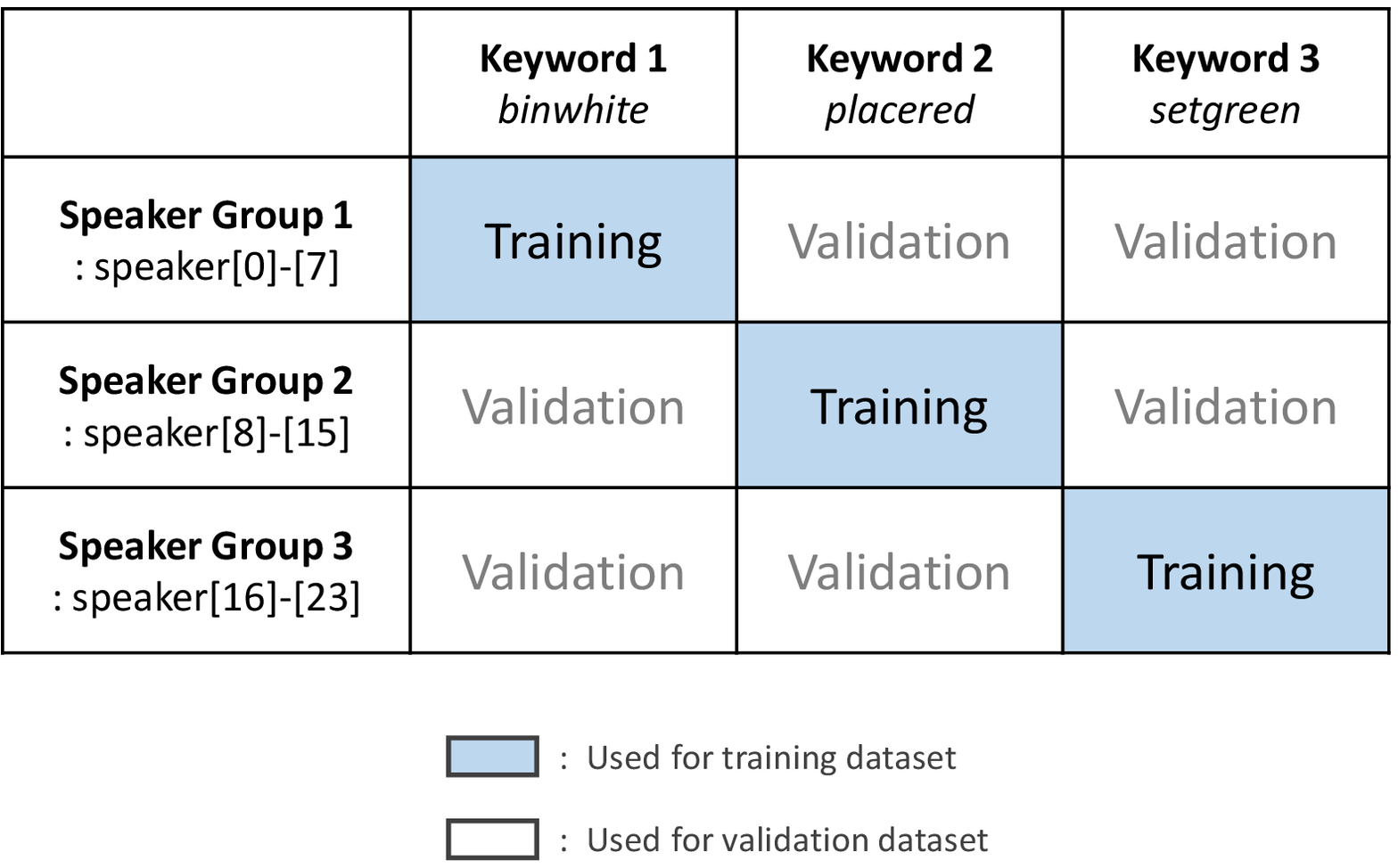,height=4.7cm}}
	\caption{An example of chosen keywords and speakers for the training and validation set when $N$=3.
	} \label{fig:dataset_split}
\end{figure}

In fine-tuning the baseline SE network, we used the CHiME 2013 database \cite{vincent2013second} which was created for the 2nd speech separation and recognition challenge with two tracks. In this experiment, we chose the track1 database consisting of keyword utterances from 34 speakers. Each utterance in CHiME database consists of a sequence of six words: command, color, preposition, letter, number, and adverb. For the short-keyword SV experiment, we used only the first two words by segmenting the utterances using the word boundary labels provided by the database. With this segmentation, we obtained the utterances of 16 different keywords: 4 types of commands ({\it bin}, {\it place}, {\it set}, {\it lay}) followed by 4 types of colors ({\it white}, {\it red}, {\it green}, {\it blue}). We chose two, three, and four keywords among them to construct the ASR network classifying $N$ = 2, 3, and 4 keywords. In this experiment setup, the number of keywords to classify is small, and thus we used one-layer DNN for the ASR network given the 128-dimensional input embedding vector. The database was split into the training-validation set of 24 speakers and the evaluation set of 10 speakers without any speaker overlap. In the training, we chose only one keyword data for a speaker, and the other keywords' data were used for the validation. For example, Fig. \ref{fig:dataset_split} shows the chosen keywords and speakers for the training and validation set when $N=3$.

The networks were optimized with stochastic gradient descent with momentum of 0.9 and weight decay of 0.00001 for the baseline training and fine-tuning stages. Also, we applied different values of $\gamma$ to see the effect of adversarial factor.

\subsection{Evaluation Result}

With the fine-tuned SE network, we performed the speaker verification using the test data: for each speaker, 5 utterances were used to obtain the enrollment vectors ${\bf X}_{ref}$, and remaining utterances were used as verification. For the text-independent SV experiment, we chose all 5 enrollment utterances from the same keyword while the verification utterances contain both the same keyword (target keyword: TK) and different keywords (non-target keyword: NTK) from the enrollment. Similar experiment setup can be found in \cite{wan2018generalized} where two keywords ('OK-Google' and 'Hey-Google') were differently used in the enrollment and verification data.

\begin{table}[]
	\begin{center}
			\caption{The EER (\%) of four different SV models (GMM-BUM, DeepRes-Base, DeepRes-GE2E and DeepRes-TriKwdAdv) with different number of keywords ($N=2, 3$, and $4$).}
			\label{table:comparison}
		\begin{tabular}{cc|ccc}
			$N$                 & model & TK & NTK & Avg. \\ \hline
			\multirow{3}{*}{2} 	& GMM-UBM   & {\bf 1.76}  & 20.85 & 11.31           \\
			& DeepRes-Base  & 2.72 & 9.95 & 6.34          \\
			& DeepRes-GE2E  &  1.83 & 8.75 & 5.29          \\
			& DeepRes-TriKwdAdv 	& 1.96 & {\bf 4.76} & {\bf 3.36}           \\ \hline
			\multirow{3}{*}{3} 	& GMM-UBM   & 2.48 & 17.78 & 10.13           \\
			& DeepRes-Base  & 2.68 & 8.10 & 5.39          \\
			& DeepRes-GE2E 	& 2.23 & 8.16 & 5.20           \\ 
			& DeepRes-TriKwdAdv 	& {\bf 2.16} & {\bf 5.15} & {\bf 3.65}           \\ \hline
			\multirow{3}{*}{4} 	& GMM-UBM   & 2.55 & 17.68 & 10.12           \\
			& DeepRes-Base  & 3.10 & 7.45 & 5.27          \\
			&  DeepRes-GE2E 	& {\bf 1.90} & 6.44 & 4.17          \\
			&  DeepRes-TriKwdAdv 	& 2.19 & {\bf 5.32} & {\bf 3.76}          \\
		\end{tabular}
	\end{center}
	
\end{table}


With this experiment setup, we evaluated four different SV models: GMM-UBM and the proposed SV frameworks without fine-tuning (DeepRes-Base), fine-tuned with GE2E-loss \cite{wan2018generalized} (DeepRes-GE2E), and fine-tuned with triplet and keyword-adversarial loss (DeepRes-TriKwdAdv). In Table. \ref{table:comparison}, the performances of four SV models with different numbers of keywords ($N$ = 2, 3, and 4) are summarized. For the performance metric, we used the EER of TK and NTK: TK is the verification performance when testing the same keyword with the enrollment keyword, and NTK is that when testing the different keyword from the enrollment keyword. Thus, better performance of NTK shows less text-dependency in speaker verification. As expected, the table shows that the EER of NTK is higher than that of TK. Especially, the EER difference between TK and NTK in GMM-UBM is quite high since the model is trained with only the target keyword, i.e. text-dependent model. The DeepRes-Base also shows big EER difference between TK and NTK. The GE2E loss which is proposed for the text-independent SV \cite{wan2018generalized} improved the EER over the baseline model (DeepRes-Base), but only marginal improvements were observed. With the proposed model, DeepRes-TriKwdAdv, we obtained considerable improvements over the GMM-UBM and also GE2E in evaluating NTK. The Avg. in the table is the mean of TK EER and NTK EER. Overall, the DeepRes-TriKwdAdv obtained the best performance for all cases: $N$=2, 3, and 4.


In Table. \ref{table:triplet}, the performances of the proposed SV framework with different numbers of keywords ($N$ = 2, 3, and 4) and adversarial factors ($\gamma$ = 0.0, 0.2, and 0.4) are summarized. The factor $\gamma=0.0$ means no adversarial training was applied: only triplet loss was used to obtain the model. When $\gamma=0.0$, the keyword accuracies (Kwd Acc.) of the ASR network show high performance for all cases of $N=2, 3$, and $4$. With increasing of $\gamma$, the Kwd Acc. decreases and also the EER of the NTK decreases. These results show that the keyword-adversarial training reduces the keyword dependency of the speaker embedding vectors. For all cases ($N$ = 2, 3, and 4), the best performance was obtained when $\gamma=0.4$. 


\begin{table}[]
		
	\begin{center}
		\caption{The performance of the proposed SV framework with different number of keywords ($N$ = 2, 3, and 4) and adversarial factors ($\gamma$ = 0.0, 0.2, and 0.4). We use the performance metric, EER (\%) of TK and NTK. The Avg. is the mean of TK and NTK, and Kwd Acc is the accuracy of the ASR network.}
		\label{table:triplet}
		\begin{tabular}{cc|cccc}
			$N$                  & $\gamma$ & TK & NTK & Avg. & Kwd Acc. \\ \hline
			\multirow{4}{*}{2} 	& 0.0 & 2.46  & 9.89  & 6.17  & 98.61  \\
			& 0.2 & 2.33 & 6.22 & 4.27 & 47.22 \\
			& 0.4 & 1.96 & 4.76 & 3.36 & 50.38 \\ \hline 
			\multirow{4}{*}{3} 	& 0.0 & 2.27 & 7.79  & 5.03 &  98.61 \\
			& 0.2 & 2.59 & 6.45 & 4.52 & 36.11 \\
			& 0.4 & 2.16 & 5.15 & 3.65 & 34.72 \\ \hline 
			\multirow{4}{*}{4} 	& 0.0 & 2.84 & 7.49 & 5.16 & 95.83  \\
			& 0.2 & 2.32 & 6.35 & 4.34 & 26.39 \\
			& 0.4 & 2.19 & 5.32 & 3.76 & 27.78 \\ \hline 
		\end{tabular}
	\end{center}
	
\end{table}

\section{Conclusion and Future Work}

In this paper, we presented an end-to-end text-independent speaker verification framework by
considering the SE network and ASR network jointly. The SE network takes the raw waveform and outputs the embedding vector which distinguishes the speaker characteristics of the input utterance. In this research, we used a number of residual blocks, convolution layer, and attention for the SE network. To obtain more discriminative and text-independent speaker embedding vectors,
we consider both the triplet loss and ASR network in training the SE network. The triplet loss maximizes the similarity between the embedding vectors of the same speaker and also minimizes that between those of different speakers. The ASR network is trained to recognize the phonetic context of the input, and using the adversarial gradient from the ASR network, the text-dependency of the embedding vectors can be reduced. In this research, we used one-layer DNN for ASR to classify $N$=2, 3, and 4 isolated keywords. 
We evaluated our SV framework using the LibriSpeech and CHiME database. With the LibriSpeech database, we trained the baseline SE network. And, then we fine-tuned the baseline model using the CHiME database by applying the proposed triplet and ASR-adversarial loss. For the short-keyword SV evaluation, we segmented the first two words of the utterances in CHiME database. In the experiments, we compared the proposed SV framework (DeepRes-TriKwdAdv) with the other 
algorithms (GMM-UBM, DeepRes-Base, and DeepRes-GE2E) which do not utilize the ASR network for the text-independent SV. In all experiments, DeepRes-TriKwdAdv outperformed the other SV models in EER. Especially, in the evaluation of NTK, the DeepRes-TriKwdAdv shows a considerable improvement over the DeepRes-Base. In this research, we set the ASR network as an isolated word classifier. For the further work, we will continue this work for more general cases: train the SE network with a general speech recognizer.


\bibliographystyle{IEEEtran}
\bibliography{user_verification}


\end{document}